\def\pr{{\it Phys. Rev. }} 
\def\jap{{\it J. Appl. Phys. }}
\def\prl{{\it Phys. Rev. Lett. }}

\documentstyle[preprint,prb,aps,epsf]{revtex}

\begin{document}
\draft
\title{Simple Mechanism for a Positive Exchange Bias} 
\author{T. M. Hong}
\address{Department of Physics, National Tsing Hua University,
Hsinchu, Taiwan 30043, R.O.C.}
\address{Department of Physics, University of California - San Diego,
La Jolla, CA 92093-0319}
\date{\today}
\maketitle

\begin{abstract}
We argue that the interface coupling, responsible for the positive exchange
bias ($H_E$) observed in
ferromagnetic/compensated antiferromagnetic (FM/AF) bilayers, favors an antiferromagnetic alignment.
At low cooling field this coupling polarizes the AF spins close to
 the interface, which spin configuration persists after the sample is  
cooled below the N${\acute e}$el temperature. This pins the FM spins
as in Bean's model and gives rise to a negative $H_E$. When the cooling
field increases, it eventually dominates and polarizes the AF spins in
an opposite direction to the low field one. This results in a positive 
$H_E$. The size of $H_E$ and the crossover cooling field are estimated.  
We explain why $H_E$ is mostly positive for an AF single crystal, and 
discuss the role of interface
 roughness on the magnitude of $H_E$, and the quantum aspect of the
interface coupling.  
 
\end{abstract}


\narrowtext
In recent experiments\cite{sh} by Schuller {\it et al.} a ferromagnetic (FM) 
film is grown on a
{\it compensated} antiferromagnetic (AF) 
surface under large cooling field, and
the hysteresis loop is observed to shift along the {\it positive} side
 of the field axis. This phenomena belongs to the general category of 
exchange anisotropy, 
first  discovered
more than 40 years ago by Meiklejohn and Bean\cite{be}. However, different 
from the original observation and later theories\cite{be,ma,mal}, a compensated
(i.e., no net magnetization) surface was used and the sign of the
bias was unexpected. The compensated part\cite{sh,no} is resolved
 by a recent 
micromagnetic calculation by Koon\cite{ko}, but the sign remains only 
speculations\cite{ra}. Without knowledge of the detail structure at the 
interface or confirmation of the existence of AF domains, we try to build
a simple intuitive theory. It not only  can
explain the main features of the phenomena, but also
gives reasonable quantitative estimations. Quantum aspect of the interface
coupling is analyzed in the second half of the paper.

Experimentally the exchange bias, $H_E$,
decreases with increasing temperature and vanishes at the N${\acute e}$el
temperature\cite{review}. This shows that the coupling
of the FM spins to the ordered AF spins is crucial for the exchange bias.
Furthermore\cite{review}, the plot of $\ln H_E$ v.s. $\ln t_F$ is found to
fit nicely by 
a straight line with slope$=-1$ where $t_F$ denotes the thickness of the FM
film. This can be viewed as another support 
to concentrate on the interface coupling for the source of exchange bias. The 
interpretation is based on, if the F spins at the interface (of number $N$) are
stabilized each by an energy, $E$, due to their coupling to the AF spins,
 we need to divide the total change,
 $NE$, by the total number of FM spins in the film, $\approx Nt_F$,
when converting to the shift in the hysterisis loop. This gives 
 $H_E\approx E/t_F$ which explains the linearity and its slope in the
ln-ln plot. Interface
roughness will increase the interface area (while the total spin number remains
unchanged) and introduce an extra factor $\alpha>1$ into $H_E\approx \alpha E
/t_F$. However, when the easy axes of FM and AF are parallel,
 surface roughness may also  
introduce frustrations (see Fig. 1) which will diminish the coupling.
This does not happen when the easy axes are perpendicular since there is no
 preferred direction for any FM spin from its neighboring AF spins. 
We shall distinguish these two orientations, parallel/perpendicular easy
 axes, and
assign them separately to the negative/positive $H_E$ cases. Such a 90
degree rotation of the FM easy axis for Fe/(110)FeF$_2$ single crystal
 due to the 
AF ordering was indeed observed\cite{perp} by examining the hysteresis loops.   
That is, the easy axis of FM spins, originally in the (001) direction at
 $T=300$ K, rotates to (1$\bar 1$0) at $T=10$ K for which
 a positive $H_E$ was measured.

For our theory, the interface coupling, $\sum J_c {\vec S_F}
\cdot {\vec S_{AF}}$, is assumed to favor an antiferromagnetic alignment
with $J_c\approx J_{AF}$ (the coupling constant between AF spins).
This will be justified if, take Fe/FeF$_2$ for instance, the fluoric ions
happen to lie at the interface and mediate the coupling between
neighboring irons from either sides (the superexchange mechanism\cite{ziman}).
 However, if 
the irons across the interface build a direct chemical bond, presumably
their coupling
will of the same order and sign as $J_F$ in the bulk Fe.
But since $J_F$ is twenty times\cite{jf} stronger than $J_{AF}$, 
this ${\vec S_{AF}}$ will be locked rigidly parallel to ${\vec S_F}$
and can be treated as an extension of the ferromagnetic film. The relevant
interface will now be between this first layer and the next layer of the antiferromagnetic, which
of course favors an antiferromagnetic alignment and agrees with our assumption. 

At low cooling field for which $H_E$ is negative, the easy axes
 of FM and AF spins are assumed to be 
parallel. Using the mean field analysis,  we estimate the deviation 
from  the positive $z$-axis
 (which results in a nonzero
magnetization for AF) of each spin-up ${\vec S_{AF}}$ at
 the interface due to 
its antiferromagnetic coupling with a spin-up ${\vec S_F}$ neighbor is
of the order of $1-\tanh \big[ (J_{AF}\cdot q-J_c)/k_BT\big]$ where $q$ is
the number of nearest neighbors for each ${\vec S_{AF}}$ at the surface. At the
usual operating temperature, say $T=10$ K, the magnetization is approximately
$-2\cdot\exp\big[- (J_{AF}\cdot q-J_c)/k_BT\big] $.
To obtain the total energy change for the system, we need to multiply
 it by $J_c$ 
and $N/2$ (number of up-spin ${\vec S_{AF}}$ at the interface).
Note that this magnetization points antiparallel to the FM spins.

When the cooling field is large enough to cause a positive $H_E$,
we assume that the easy axis of FM spins rotates and becomes perpendicular to
the AF easy axis. Different from the previous case, polarization of the
AF spins is now mainly due to the cooling field and, not just those ${\vec
S_{AF}}$ at the interface but, all spins are involved.
A physical justification for making such a rotation may lie in the
 fact\cite{ziman} that
the perpendicular magnetic susceptibility of AF spins ($\approx 1/J_{AF}$) is
much larger than the parallel one at low temperatures. By canting the AF 
spins more effectively towards the field direction the system can gain more
energy from the Zeeman effect.
Note that the polarization here	points parallel to 
the external field, i.e., the easy axis of FM spins, and is opposite to that 
 caused by the interface coupling.

We can estimate the 
minimum strength of cooling field, $H_{cool}$, required to obtain
 a positive magnetization by comparing these two energy changes:
\begin{equation}
\frac{H^2_{cool}}{J_{AF}}\cdot t_{AF}N=
J_c\ e^{- (J_{AF}\cdot q-J_c)/k_BT}\cdot N
\end{equation} 
This gives $H_{cool}\approx 0.2$ T for $t_{AF}=90$ nm, the right magnitude
to cause the sign change
of $H_E$ experimentally\cite{sh}. Had the easy axes been perpendicular 
at low cooling field, RHS of the above equation would become $J^2_cN/J_{AF}$
and give too high a threshold field $H_{cool}\approx 5$ T. 
The thickness $t_{AF}$ becomes very large
for a single crystal, 
which implies an easier entrance into the positive-$H_E$ scenario.
This is again consistent with observations\cite{review} that $H_E$ is mostly positive when an
AF single crytal is used (the fact that its surface is much rougher than
in films also contributes).
The appearance of an exchange bias due to the 
locking of FM spins by the polarization is the same as in Bean's original
model\cite{be}, except that an uncompensated AF surface is not required here 
and $H_E$ can become positive when the cooling field is strong.
We do not know how the polarization survives below the N${\acute e}$el
temperature. This could be the place where possible AF domains or impurities
need to be introduced. Experimental evidence\cite{sh} for this "memory" is
found when putting samples, field cooled in 2 kOe, under 70 kOe magnetic
field at low temperatures (10K). $H_E$ is found to remain unchanged
to within 5$\%$ of the $H_{cool}=2$ kOe value.

Aside from possible instability due to finite temperature fluctuations,
the main conclusion of Koon\cite{ko} that FM orders perpendicular
to the AF easy magnetization axis was checked to be correct by Kiwi\cite{ra}
using a Monte-Carlo calculation.
We shall examine the validity of this conclusion against a  
full quantum mechanical treatment, i.e.,  we analyze the 
change of vacuum energy, $E_{vac}$, due to the virtual process of FM spins
emitting and reabsorbing AF spin waves via the interface coupling.
Suhl and Schuller\cite{suhl} have considered the special case when the FM and AF
easy axes are parallel, and found a negative $E_{vac}$. We extend their
calculations to a general angle, $\phi$, between these two easy axes
 (see Fig.2) in
order to find the most stable spin orientation. 

Quantum mechanically the interface coupling, $\sum J_c {\vec S_F}\cdot
 {\vec S_{AF}}$ where the summation runs over all sites at the interface,
can be decomposed into raising and lowering operators as
 $[S^+_F S^-_{AF}+S^-_F S^+_{AF}]/2+S^z_F S^z_{AF}$. Since the easy axis
of ${\vec {S_F}}$ is now in the $(0,\sin\phi ,\cos\phi )$ direction, 
we need to reexpress $S^z_F$ and $S^\pm_F$ in terms of the new projection
and raising/lowering operators: 
\begin{eqnarray}
\nonumber
P^z&\equiv& S^y_F\sin\phi+S^z_F\cos\phi\\
P^\pm&\equiv& S^x_F\pm i\big( S^y_F\cos \phi -S^z_F\sin\phi \big).
\end{eqnarray}
In the mean time follow
the standard spin wave derivation\cite{ziman} in rewriting the AF
spin operators in terms of 
boson operators $a^+$ and $a$ which create and destroy spin deviations,
\begin{eqnarray}
\nonumber
S_l^x&\approx& \sqrt{\frac{s}{2}}\ \big( a_l+a^+_l\big),\\
S_l^y&\approx& -i\sigma_l \sqrt{\frac{s}{2}}\ \big( a_l-a^+_l\big),\\
\nonumber
S_l^z&=&\sigma_l \big( s-a^+_l a_l\big),
\label{bo}
\end{eqnarray}
where $l$ is the site label and  $\sigma_l=1/-1$ at the spin-up/down
${\vec S_{AF}}$ site,
and $s/S$ is the size of the AF/FM spin. Since there is
no confusion now between the different spin notations, we shall drop the
subscripts $F$ and $AF$ from now on.

The interface coupling becomes $ J_c$ times 
\begin{eqnarray}
\nonumber
\sum_u\Bigg[P^z_u\cos\phi-\frac{P^+_u-P^-_u}{2i}\sin\phi&\Bigg]&\ \big(
s-a^+_u a_u\big)
-\sum_d\Bigg[P^z_d\cos\phi-\frac{P^+_d-P^-_d}{2i}\sin\phi\Bigg]\ \big(
s-a^+_d a_d\big)\\
\nonumber
+\sqrt{\frac{s}{2}}\sum_u\Bigg\{
&\Big [&\frac{P^+_u+P^-_u}{2}+iP^z_u\sin\phi+\frac{P^+_u-P^-_u}{2}\cos\phi
\Big ]a^+_u\\
\nonumber
+&\Big[&\frac{P^+_u+P^-_u}{2}-iP^z_u\sin\phi-\frac{P^+_u-P^-_u}{2}\cos\phi
\Big ]a_u\Bigg\}\\
\nonumber
+\sqrt{\frac{s}{2}}\sum_d\Bigg\{
&\Big [&\frac{P^+_d+P^-_d}{2}+iP^z_d\sin\phi+\frac{P^+_d-P^-_d}{2}\cos\phi
\Big ]a_d\\
+&\Big [&\frac{P^+_d+P^-_d}{2}-iP^z_d\sin\phi-\frac{P^+_d-P^-_d}{2}\cos\phi
\Big ]a^+_d\Bigg\}
\label{hc}
\end{eqnarray}
where the subscript $u/d$ denotes the site with or  neighboring  a
down/up AF spin (see Fig.2).
The fact that 
the ordering temperature of FM ($\approx 770$ K) is
much higher than that of AF ($\approx 78$ K) allows us to assume
that the FM spins are very much rigid 
while the AF spins deviate. In the ground state, the brackets $s-a^+_u a_u$
and $s-a^+_d a_d$ have the same values, and the first and the second terms 
of Eq.~(\ref{hc}) cancel. Now expressing $a^+_{u,d}$ and $a_{u,d}$ in terms
of their Fourier conjugates $a^+_k$ and $a_k$, and then Bogoliubov
 transforming\cite{ziman} to the AF spin wave operators $b_k$ and $b^+_k$:
\begin{eqnarray}
b_k&=&a_k\cosh\theta_k +c^+_{-k}\sinh\theta_k,\quad
 b^+_k=a^+_k\cosh\theta_k +c_{-k}\sinh\theta_k,\\
b_{-k}&=&a^+_k\sinh\theta_k +c_{-k}\cosh\theta_k,\quad 
b^+_{-k}=a_k\sinh\theta_k +c^+_{-k}\cosh\theta_k.
\end{eqnarray}
Eq.~(\ref{hc}) becomes 
\begin{equation}
J_c\sqrt{\frac{s}{2N}}\sum_k \Big( q_k b^+_k+q^+_k b_k\Big)
\label{coup}
\end{equation}
where 
\begin{eqnarray}
\nonumber
q_k\equiv \sum_{u}\Bigg\{&\Big[&\frac{P^+_u+P^-_u}{2}+iP^z_u\sin\phi+
\frac{P^+_u-P^-_u}{2}\cos\phi\Big]\cosh\theta_k\cdot e^{iku}\\
\nonumber
+&\Big[&\frac{P^+_u+P^-_u}{2}-iP^z_u\sin\phi-
\frac{P^+_u-P^-_u}{2}\cos\phi\Big]\sinh\theta_k\cdot e^{-iku}\Bigg\}\\
\nonumber
+\sum_{d}\Bigg\{&\Big[&\frac{P^+_d+P^-_d}{2}+iP^z_d\sin\phi+
\frac{P^+_d-P^-_d}{2}\cos\phi\Big]\sinh\theta_k\cdot e^{-ikd}\\
+&\Big[&\frac{P^+_d+P^-_d}{2}-iP^z_d\sin\phi-
\frac{P^+_d-P^-_d}{2}\cos\phi\Big]\cosh\theta_k\cdot e^{ikd}\Bigg\}.
\label{hc2}
\end{eqnarray} 

We can redefine $b_k$ and $b^+_k$ to eliminate the linear terms in 
Eq.~(\ref{coup}). This shifts the 
vacuum energy of the antiferromagnetic Hamiltonian, $\sum_k \omega_k
b^+_k b_k$, by $E_{vac}=-\big(sJ_c^2/2N\big)\sum_{k}
\frac{q^*_k q_k}{\omega_k}$.
The summation can be written out as 
\begin{eqnarray}
\nonumber
\sum_{k}\frac{1}{\omega_k}
\Bigg\{&\sum_{u}&\frac{P^+_uP^-_u}{4}\Big[(1+\cos\phi )^2\cosh^2\theta_k +
(1-\cos\phi )^2\sinh^2\theta_k+
\sin^2\phi\ \sinh 2\theta_k\cos 2ku\Big]\\
\nonumber
+&\sum_{d}&\frac{P^+_dP^-_d}{4}\Big[(1+\cos\phi )^2\sinh^2\theta_k +
(1-\cos\phi )^2\cosh^2\theta_k+
\sin^2\phi\ \sinh 2\theta_k\cos 2kd\Big]\\
+\sin^2\phi\Big[&\sum_u& \big(P^z_u\big)^2\Big(\cosh 2\theta_k -\sinh 2\theta_k
\cos 2ku\Big)
+\sum_d\big (P^z_d\big)^2\Big(\cosh 2\theta_k -\sinh 2\theta_k
\cos 2kd\Big)\Big]\Bigg\}
\label{vac}
\end{eqnarray}
where cross terms $P^\pm_uP^\pm_d$ and products of two raising or lowering
operators have been neglected since we are averaging over the ground state.
 According to the completeness relation, 
 $\sum_u\cos 2ku+\sum_d\cos 2kd$ is only nonzero when $k=0$ or $\pi$ 
(setting the lattice constant to be unity). $k=0$ mode is neglected
on physical ground since it 
involves translationally moving the whole sample and is not what we expect. 
Substituting the groundstate expectation values of $\langle P^+_{u/d}
P^-_{u/d}\rangle=2S$
and $\langle \big(P^z_{u/d}\big)^2\rangle =S^2$ into Eq.~(\ref{vac}) gives 
\begin{equation}
E_{vac}=-\frac{sJ_c^2}{2}\Bigg\{ \Big[ S\big(1+\cos^2\phi \big)+S^2\sin^2\phi\Big]
\sum_{k}\frac{\cosh 2\theta_k}{\omega_k}
+\big(\frac{S}{2}-S^2\big)\sin^2\phi
\ \frac{\sinh 2\theta_\pi}{\omega_\pi}\Bigg\}
\label{result}
\end{equation}
The summation of $\cosh 2\theta /\omega$ is of the order of $\ln
\big(J_{AF}/H_A\big)/J_{AF}$,
while $\sinh 2\theta_\pi/\omega_\pi\approx 1/H_A$ 
where $H_A$ is the
anisotropic field for the AF spins. Normally, $H_A$ is of the order of
a few hundred gauss and much smaller 
than $J_{AF}$, and so
we expect the second term in Eq.~(\ref{result}) to dominate as long as
$\phi\neq 0$ or $\pi$. When 
$S$ is bigger than $1/2$ (experimentally $S=2$), $E_{vac}$
becomes positive. This means that the system is more stable when the
FM/AF easy axes are parallel, compared to being perpendicular, which
is opposite to the conclusion from micromagnetic
calculations\cite{ko}. Of course, the inclusion of anisotropy field on both
sides, finite temperature, and couplings in further layers are necessary 
to determine the final preference. But
at least the above calculations show that the quantum nature of the interface
coupling, neglected in former classical treatments\cite{be,ma,mal,ko,ra}, 
may reverse their conclusions and should be properly taken into account.

In conclusion, we have presented a simple mechanism to explain why the 
sign of exchange bias, $H_E$, changes from negative to positive at high cooling
field for the
FM/compensated AF bilayers. We believe that the negative/positive bias is 
realized when 
the easy axes of FM and AF are parallel/perpendicular. And the time reversal
symmetry is broken by the polarization induced by the
 interface coupling/cooling 
field respectively. We can explain why the surface roughness enhances $H_E$
when it is positive, while diminishes its magnitude when negative.
We also estimate the right size of $H_E$ and the crossover cooling field.
Quantum nature of the interface coupling is shown to give an opposite
preference of spin alignments to former classical treatments.

This work is supported by the NSC of Taiwan under Contract No. 
87-2112-M-007-008. I want to thank Professor Ivan Schuller for
introducing and sharing with me his enthusiasm to this problem. 
Suggestions from Professor Harry Suhl have been critical and relevant as usual.
I also benefited from discussions with my host, Professor Lu Sham, 
Professor Scot Renn, Professor Gerd Czycholl, Professor Miguel Kiwi,
Dr. Norbert Linder and Dr. Josop Nogues.

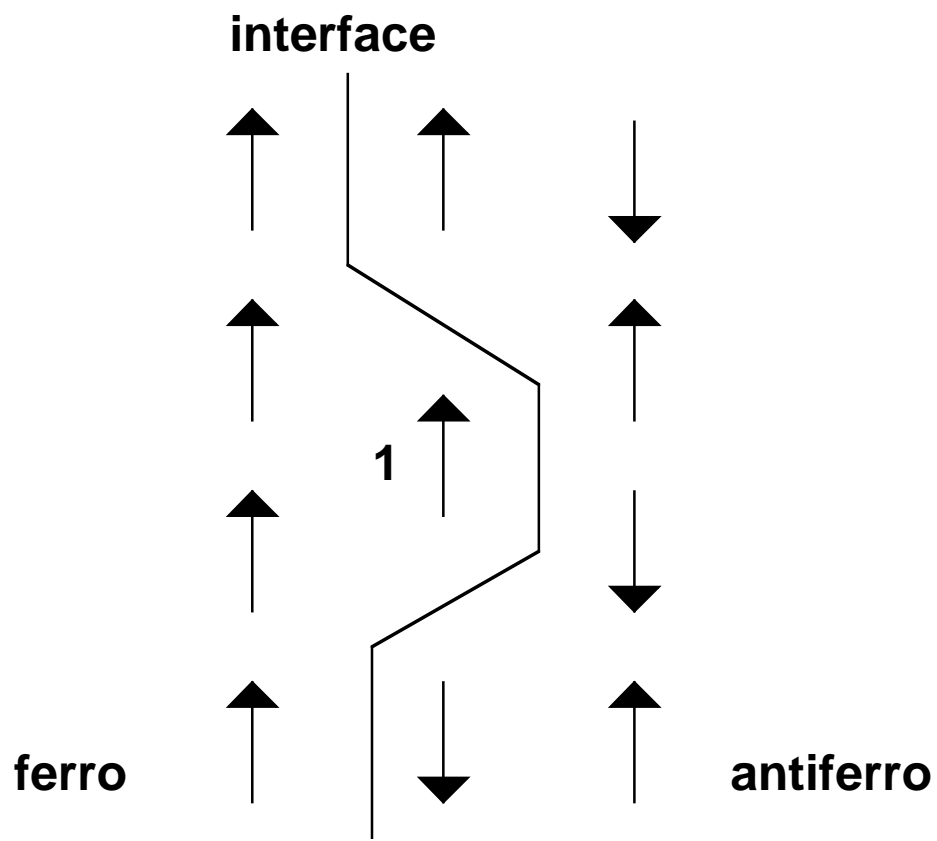
\begin{figure}
\caption{ Roughness introduces frustration to the ferromagnetic spin 1
when its neighbouring spins are in opposite directions.}
\end{figure}

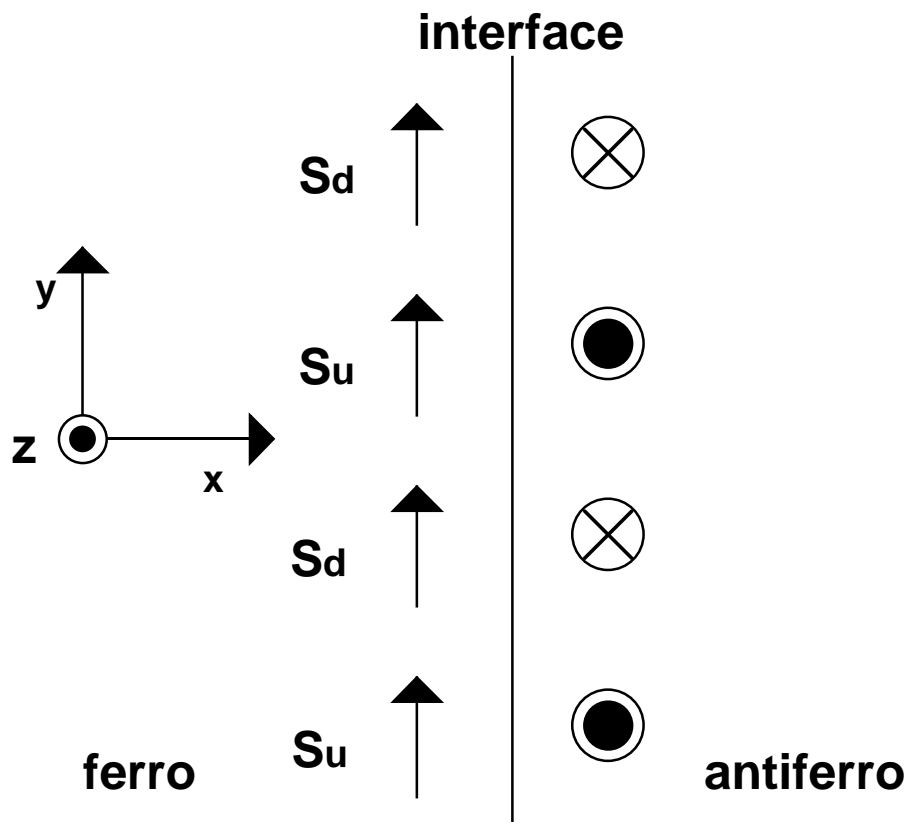
\begin{figure}
\caption{Interface between the ferromagnetic and fully compensated
 antiferromagnetic layers modeled in the text. The cross/dot symbol denotes
down/up spins. The z-axis is defined to be along the AF easy axis, and FM spins point in $(0,\sin\phi ,\cos\phi )$ direction.}
\end{figure}


\begin{references}
\bibitem{jf} We estimate that $J_F\approx 260$ K and $J_{AF}\approx 20$ K
	from the Curie and the N${\acute e}$el temperatures, 1040 and 78 K.
\end{references}
\end{document}